\documentclass[aps, prx, twocolumn, superscriptaddress, longbibliography]{revtex4-2}

\usepackage{graphicx}
\usepackage{dcolumn}
\usepackage{bm}
\usepackage{hyperref}
\usepackage{color}
\usepackage{glossaries}
\usepackage{sidecap} 
\usepackage{siunitx}
\usepackage{orcidlink}

\newacronym{DFT}{DFT}{density functional theory}
\newacronym{RLU}{RLU}{reciprocal lattice units}
\newacronym{WSM}{WSM}{Weyl semi-metals}
\newacronym{ZFC}{ZFC}{zero-field-cooled}
\newacronym{FC}{FC}{field-cooled}
\newacronym{SOC}{SOC}{Spin-orbit coupling}
\DeclareSIUnit\angstrom{\text {Å}}

\begin{document}

\title{Field-driven commensurate magnetism in EuAuSb}
\title{Frustrated helical magnetic ordering in semimetallic EuAuSb}
\title{Field-driven incommensurate-to-commensurate helical magnetic order in EuAuSb}
\title{EuAuSb: An incommensurate helical variation on an altermagnet}
\title{EuAuSb: An odd-parity helical variation on altermagnetism}

\author{J. Sears\orcidlink{0000-0001-6524-8953}}
\email{jsears@bnl.gov}
\affiliation{Condensed Matter Physics and Materials Science Division, Brookhaven National Laboratory, Upton, New York 11973-5000, USA}

\author{Juntao Yao\orcidlink{0000-0002-4339-1027}}
\affiliation{Condensed Matter Physics and Materials Science Division, Brookhaven National Laboratory, Upton, New York 11973-5000, USA}
\affiliation{Department of Materials Science and Chemical Engineering, Stony Brook University, Stony Brook, New York 11794-3800, USA}

\author{Zhixiang Hu\orcidlink{0000-0001-7925-2662}}
    \affiliation{Department of Materials Science and Chemical Engineering, Stony Brook University, Stony Brook, New York 11794-3800, USA}

\author{Wei Tian\orcidlink{0000-0001-7735-3187}}
\affiliation{Neutron Scattering Division, Oak Ridge National Laboratory, Oak Ridge, Tennessee 37831, USA}

\author{Niraj Aryal\orcidlink{0000-0002-0968-6809}}
\author{Weiguo Yin\orcidlink{0000-0002-4965-5329}}
\author{A. M. Tsvelik\orcidlink{0000-0002-7478-670X}}
\author{I. A. Zaliznyak\orcidlink{0000-0002-8548-7924}}
\affiliation{Condensed Matter Physics and Materials Science Division, Brookhaven National Laboratory, Upton, New York 11973-5000, USA}

\author{Qiang Li\orcidlink{0000-0002-1230-4832}}
\affiliation{Department of Physics and Astronomy, Stony Brook University, Stony Brook, New York 11794-3800, USA}
\affiliation{Condensed Matter Physics and Materials Science Division, Brookhaven National Laboratory, Upton, New York 11973-5000, USA}

\author{J. M. Tranquada\orcidlink{0000-0003-4984-8857}}
\email{jtran@bnl.gov}
\affiliation{Condensed Matter Physics and Materials Science Division, Brookhaven National Laboratory, Upton, New York 11973-5000, USA}

\date{\today}

\begin{abstract}

EuAuSb is a triangular-lattice Dirac semimetal in which a topological Hall effect has been observed to develop in association with a magnetically-ordered phase. Our single-crystal neutron diffraction measurements have identified an incommensurate helical order in which individual ferromagnetic Eu$^{2+}$ layers rotate in-plane by $\sim$120$^{\circ}$ from one layer to the next. An in-plane magnetic field distorts the incommensurate order, eventually leading to a first order transition to a state that is approximately commensurate and that is continuously polarized as the bulk magnetization approaches saturation.  From an analysis of the magnetic diffraction intensities vs.\ field, we find evidence for a dip in the ordered in-plane moment at the same field where the topological Hall effect is a maximum, and we propose that this is due to field-induced quantum spin fluctuations.  Our electronic structure calculations yield exchange constants compatible with the helical order and show that the bands near the Fermi level lose their spin degeneracy via a mechanism similar to that in the collinear altermagnets. We find that, unlike the even symmetry seen in the altermagnets, the spin-splitting in EuAuSb has odd-wave symmetry similar to that recently found in a number of coplanar magnetic materials.


\end{abstract}

\maketitle

\section{Introduction}

Materials with spin-polarized electronic states have great potential for spintronics applications \cite{smej18}. In a Dirac semimetal, conversion to a Weyl semimetal \cite{Armitage2018_weyl_dirac, Lv2021_perspective} with an energy splitting between degenerate spin states can be achieved by combining spin-orbit coupling and broken inversion symmetry, as in materials such as TaAs \cite{xu15b,lv15,weng15,huan15}, or  by breaking time-reversal symmetry through suitable magnetic ordering \cite{Bernevig2022_prospects}, as in ferromagnetic Co$_3$Sn$_2$S$_2$ \cite{yin19,liu19,mora19}.  Given the benefits of avoiding stray fields from ferromagnetism \cite{smej18,naka15}, there has been recent excitement about collinear compensated magnetic order defined as altermagnetism \cite{smej22a,smej22b,mazi22}, which has spin-split bands over significant ranges of reciprocal space. While much of this work has focused on the collinear systems originally termed altermagnets, these ideas have also been generalized to noncollinear magnetic structures \cite{cheo24, rada24tensorial, xiao24, rada25, chen24, jian24}. In this paper, we experimentally explore one such compensated noncollinear order that yields nondegenerate electronic states.

\begin{figure}
\includegraphics[width=1\columnwidth]{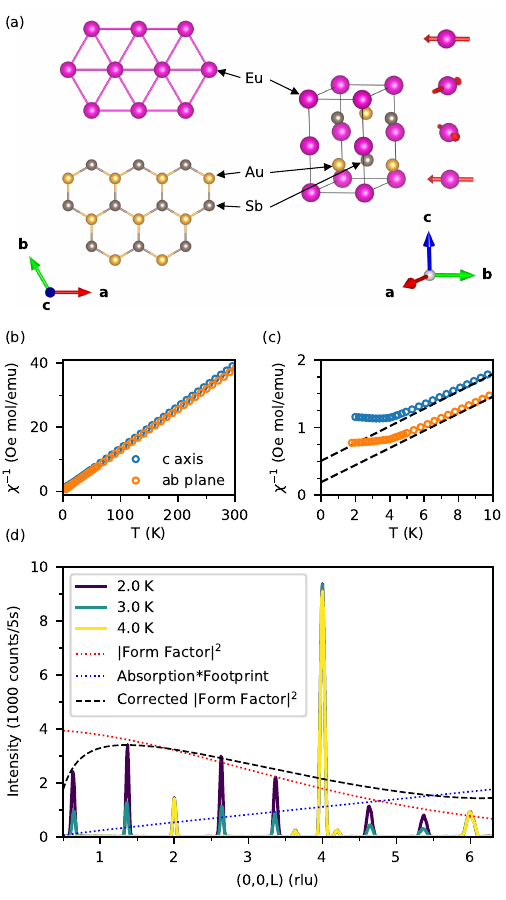}
\caption{(a) Crystal stucture of EuAuSb with space group $P6_3/mmc$, and magnetic moment directions for the Eu planes. The drawing was produced using VESTA \cite{Momma2011}. (b) Inverse of magnetic susceptibility for the $c$-axis and in-plane directions, showing linear Curie-Weiss behavior. (c) Zoom in of low temperature region of the inverse magnetic susceptibility, showing the change in slope below the magnetic ordering transition of 3.9~K. The Curie-Weiss fit to the high temperature data is shown by the black dashed lines. (d) Temperature dependence of $(0,0,L)$ scans, showing appearance of additional magnetic peaks below the magnetic ordering temperature. Bragg peaks at $(0,0,2)$, $(0,0,4)$ and $(0,0,6)$ positions are present at all temperatures. The dashed black line shows the Eu$^{2+}$ form factor, corrected for absorption, footprint, and Lorentz factor.  The weak peaks at $L=3.65$ and 4.21 rlu are from Al powder rings.}
\label{fig1}
\end{figure}

In particular, we examine a member of the family Eu$MX$, with $M=$ Cu, Ag, and Au, and $X=$ P, As, and Sb.  All members share the same hexagonal structure shown in Fig.~\ref{fig1}(a) \cite{tomu81,pott00,mishra2011structure}, in which triangular-lattice layers of Eu alternate with $MX$ layers.  The paramagnetic moment per Eu$^{2+}$ ion extracted from the temperature dependence of the magnetic susceptibility is in the range of 94\%\ to 101\%\ of the effective moment, 7.94~$\mu_{\rm B}$, expected for spin-only magnetism of the half-filled $4f^7$ shell of Eu$^{2+}$ with $J=S=7/2$ \cite{tomu81,mishra2011structure}. 

The topological Hall effect has been observed in \mbox{EuAgAs} \cite{laha21}, EuAuSb \cite{ram2024magnetotransport,roy2024chiral}, and EuCuAs \cite{roychowdhury2023interplay}.  The effect is observed in the range of temperatures and magnetic fields for which magnetization data suggest the presence of magnetic order.  There has been speculation about the nature of the magnetic order in EuAgAs and EuAuSb \cite{laha21,jin21,ram2024magnetotransport,roy2024chiral}, but no direct measurements.  In contrast, neutron diffraction studies of EuCuAs have   
identified magnetic order consisting of ferromagnetically-ordered planes of Eu atoms, where the net moment direction rotates from one plane to the next by 90$^\circ$, with a period of 4 layers, corresponding to 2 unit cells \cite{roychowdhury2023interplay,soh2024weyl}; related magnetic order has been detected in EuCuSb \cite{takahashi2020competing}. The results of modeling for EuCuAs suggest that the helical magnetic order should stabilize Weyl points in the band structure \cite{soh2024weyl}.  In the case of EuAuSb, the presence of field-induced Weyl states was inferred from observations \cite{roy2024chiral} of the chiral magnetic effect \cite{Li_2016} in longitudinal magnetoresistance at temperatures well above the magnetic ordering transition.  Hence, one may expect interesting behaviors that depend on the nature of the magnetic order.

We have therefore carried out neutron diffraction measurements on a single crystal of EuAuSb to determine its magnetic ground state. Our results indicate that the magnetic order is helical, similar to that seen in the isostructural compounds, but with a different propagation vector along the $c$ axis. The ordering wave vector was found to be $\sim(0,0,\frac{2}{3})$ with spins parallel to the planes, consistent with a rotation of the moments by $\sim$120$^{\circ}$ from one plane to the next, corresponding to a magnetic unit cell containing three Eu layers. At low temperature and zero field, the order is distinctly incommensurate, with the incommensurability decreasing as the spin spiral distorts \cite{zali95} in an applied in-plane field; as the field approaches 1~T, there is a sharp and hysteretic (first-order) transition that leads to an approximately commensurate phase. With increasing field the moments continue to cant towards the field direction, increasing the ferromagnetically ordered component as the material becomes fully polarized. We find that at intermediate fields the total observed ordered moment is suppressed, which we suggest may be due to quantum fluctuations.

Lastly, our electronic structure calculations show that the states in the vicinity of the Fermi level experience a spin-dependent energy splitting even in zero applied field.  Given that the planar-spiral magnetism that causes the spin splitting is fully compensated, this corresponds to a generalized form of altermagnetism \cite{cheo24, rada24tensorial, xiao24, rada25, chen24, jian24, smej22b}. We find that the symmetry of the spin-splitting is odd under inversion of reciprocal space, which contrasts with the even symmetry of the collinear altermagnets. Odd symmetry has been observed in a number of recently investigated coplanar magnets \cite{hellenes2024pwavemagnets, pari25, song2025pwaveswitch, kudasov2024helical, cuono2023_eucompounds}, and we identify EuAuSb as a novel member of this class of odd-wave spin-split materials.

The rest of the paper is organized as follows.  In Sec.~II, we describe our experimental and computational methods.  The results are presented and interpreted in Sec.~III.  We end with discussion in Sec.~IV and conclusions in Sec.~V.

\section{Methods}

\subsection{Crystal growth and characterization}

Single-crystal samples of EuAuSb were synthesized by the self-flux method. Eu, Au and Sb were mixed in the molar ratio of 1:1:5, loaded in an alumina crucible and then sealed in a quartz tube under vacuum. The mixture was heated from room temperature to 1000$^{\circ}$C at a rate of 100$^{\circ}$C/h and maintained at this temperature for 12 hours. The furnace was then cooled to 650$^{\circ}$C at a rate of 2$^{\circ}$C/h and the resulting crystals were separated from the Sb flux by centrifuge. Some of these single crystals were ground and measured on a tabletop
x-ray diffractometer Miniflex+ with a Cu $K_\alpha$ source. The data were refined with the software Rietica, confirming that the structure is the expected EuAuSb phase with $P6_3/mmc$ space group (as well as some residual Sb flux). The lattice parameters were found to be $a=b=\SI{4.6740(3)}{\angstrom}$ and $c=\SI{8.5293(7)}{\angstrom}$, consistent with the published crystal structure.

The samples used for the magnetization measurements were polished to remove residual flux. The magnetic properties were measured in a Quantum Design Magnetic Properties Measurement System (MPMS). 

\subsection{Neutron diffraction}

The crystal used for the neutron scattering measurement was a hexagonal plate of dimensions of $9\times6\times0.5$ mm with a mass of $\sim$0.2~g. The crystal was sanded to remove polycrystalline material and Sb flux from the surface, and to provide a large, relatively flat surface suitable for scattering. The crystal was partially wrapped in Al foil, attached to an Al plate with Al wire and oriented using x-ray Laue diffraction. The neutron diffraction experiment was carried out at the VERITAS (HB-1A) instrument at the High Flux Isotope Reactor at Oak Ridge National Laboratory. The crystal and mount were attached to the sample stick for the 6~T vertical-field magnet with a pumped-$^4$He cryostat; the base was masked with Gd$_2$O$_3$ tape. The crystal was oriented with $(H,0,L)$ in the horizontal scattering plane and the $[-K,2K,0]$ direction along the vertical magnetic field direction. Neutron diffraction scans were performed along ${\bf Q}=(0,0,L)$, where {\bf Q} was perpendicular to the surface and $L$ is in reciprocal lattice units (rlu) of $2\pi/c$.  Because of the strong neutron absorption by Eu, we were not able to detect diffraction in any other directions.

\begin{figure*}
\includegraphics[width=2\columnwidth]{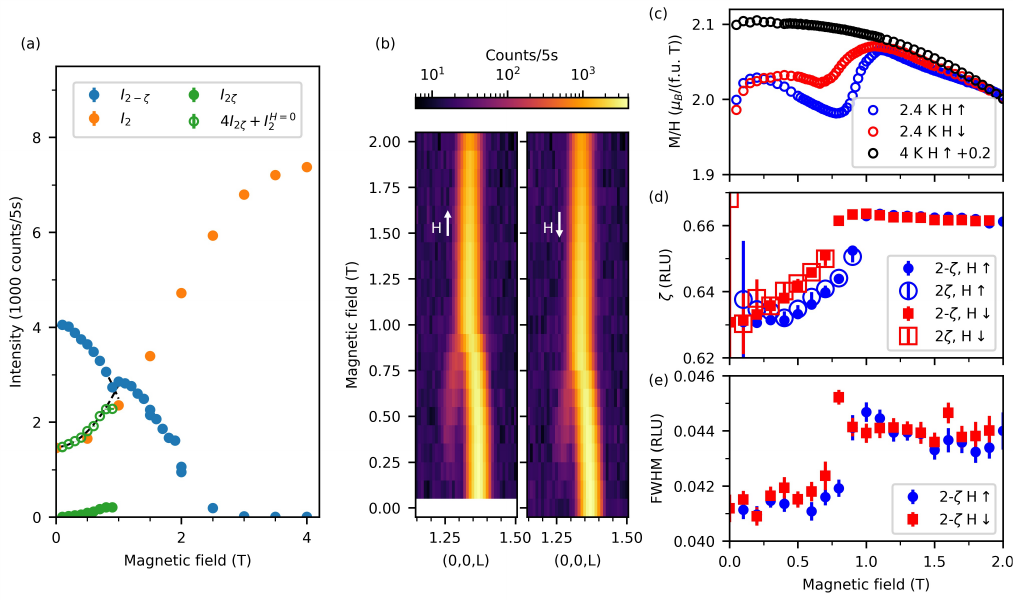}
\caption{ (a) Magnetic field dependence of the intensities for the Bragg ($I_2$) and magnetic first and second harmonic peaks ($I_{2-\zeta}$ and $I_{2\zeta}$ respectively). The black dashed lines are fits, showing that in the low field range, the first and second harmonic peak intensities have the expected quadratic field dependence. (b) Field dependence of the magnetic signal in the vicinity of the $(0,0,2-\zeta)$ magnetic peak, showing an incommensurate to commensurate transition at $\sim$1~T. The weaker feature at lower $L$ is the second harmonic. Data was collected at 1.6~K. (c) Magnetization divided by magnetic field as a function of field, showing a transition at $\sim0.9$~T. The 4~K data is offset by a constant of 0.2~$\mu_B$/f.u./T to simplify comparison with the 2.4~K data. (d), and (e) Fitted peak positions and FWHM from the fits of the 1.6~K data shown in panel (a).   }
\label{fig2}
\end{figure*}

\subsection{Computational details}

To evaluate the electronic structure,
density-functional theory (DFT) calculations were performed using the Vienna Ab Initio Simulation Package (VASP)~\cite{VASP1,VASP2}. The Perdew-Burke-Ernzerhof (PBE) exchange-correlation functional~\cite{PBE} within the generalized gradient approximation (GGA) was employed for all calculations. An energy cutoff of 500 eV was used, and a Monkhorst-Pack \textit{k}-mesh with sampling density up to 17$\times$17$\times$13 was chosen to sample the primitive Brillouin zone, ensuring the resolution of small energy differences between various magnetic phases. 
The GGA + $U_{\textrm{eff}}$ method ~\cite{LDAU_Dudarev} was used to treat the Eu-4\textit{f} orbitals, with $U_{\textrm{eff}}$ set to 6 eV~\cite{EuP_JohannesPRB2005,Kunes_Eu_JSPS2005,Eu_aryalPRB2022}. 
Calculations for the non-magnetic phase were carried out under the open-core approximation by freezing the Eu-4\textit{f} states. \gls*{SOC} was treated via the second-variational method as implemented in VASP.

\section{Results and Interpretation}

The crystal structure of EuAuSb, illustrated in Fig.~\ref{fig1}(a), is composed of triangular-lattice layers of magnetic Eu$^{2+}$ ions separated by layers of Au and Sb. Figure~\ref{fig1}(b) and (c) show the inverse of the magnetic susceptibility plotted as a function of temperature, measured with the magnetic field either along the crystallographic $c$ axis or in the $ab$ plane. The high-temperature region shown in panel (b) exhibits linear Curie-Weiss type behavior, while the zoomed-in low-temperature data in panel (c) show a small kink at $T_{\rm N}$ and deviation from linearity in the ordered phase. Features at a consistent $T_{\rm N}$ of 3.9~K were also found in resistivity and specific heat collected on multiple crystals (these results will be reported elsewhere \cite{yao25}). We note that our observed magnetic ordering temperature is similar to but distinct from previous reports on this material ($T_{\rm N}=4.1$~K \cite{mishra2011structure}, 3.5~K \cite{roy2024chiral}, 3.3~K \cite{ram2024magnetotransport}).

The modest anisotropy of the magnetic susceptibility data indicates easy-plane anisotropy of the Eu moments. Consistent with previous work \cite{tomu81,mishra2011structure,ram2024magnetotransport,roy2024chiral}, the paramagnetic moments obtained from Curie-Weiss fits (between 40~K and 300~K) to $\chi_{ab}$ and $\chi_c$ are 7.94(1)~$\mu_{\rm B}$ and 7.88(1)~$\mu_{\rm B}$, and the Curie-Weiss temperatures are $-1.5(1)$~K and $-3.9(1)$~K respectively. For both orientations, the low temperature magnetization saturates just below $g\mu_{\rm B}S = 7$~$\mu_{\rm B}$  \cite{yao25}, the value expected for magnetic Eu$^{2+}$ ions with a 4$f^7$ electron configuration. 

We collected single-crystal magnetic neutron diffraction measurements to characterize the magnetic ordering of the Eu moments below $T_{\rm N}$.  Long scans along $(0,0,L)$ are plotted in Fig.~\ref{fig1}(d), showing nuclear Bragg peaks at $L=2n=2$, 4, and 6 present at all temperatures. When the sample was cooled below $T_{\rm N}$, additional magnetic peaks appeared at positions $L=2n \pm \zeta$ where $\zeta\sim\frac{2}{3}$. The wave vector observed at the lowest temperature is slightly incommensurate, with $\zeta\approx0.63$.

The magnetic peak intensities as a function of $L$ are in reasonably good agreement with the calculated Eu$^{2+}$ form factor corrected for absorption, geometry, and Lorentz factor, as indicated by the dashed line in Fig.~\ref{fig1}(d).  The observed intensity falls off faster than the corrected form factor at high momentum transfer, likely due to inaccuracies in our absorption correction resulting from irregularities on the sample surface. We also observed a slight increase in intensity on the main Bragg peaks below $T_{\rm N}$. We note that the magnetization measurements are not consistent with a ferromagnetic component of this magnitude so this change is most likely structural in origin. The pattern of purely magnetic peaks is consistent with helical ordering, in which the moments are ferromagnetically aligned within the Eu layers and the net moment rotates from one plane to the next by $\sim$120$^{\circ}$. This type of magnetic order has been observed previously in EuCuSb \cite{takahashi2020competing} and EuCuAs \cite{soh2024weyl} but with different periodicities along the $c$ direction.

Next, we measured the effect of an in-plane magnetic field on the (0,0,2) and $(0,0,2-\zeta)$ peaks.  As shown in Fig.~\ref{fig2}(a), the general trend is that the intensity of the $(0,0,2-\zeta)$ decreases, approaching zero near 2.5~T, consistent with suppression of the spiral order, while a ferromagnetically-polarized component at $(0,0,2)$ grows and eventually approaches saturation near 4~T, similar to the uniform magnetization.

On a finer scale, interesting details appear.  As shown in Fig.~\ref{fig2}(b), scans near $L=2-\zeta=1.37$ show that, with increasing field, a second peak appears at $L=0+2\zeta$, corresponding to a second harmonic.  As the field approaches a critical value, $\mu_0H_{c1}\sim0.9$~T, $\zeta$ and the intensity of the second harmonic peak grow.  For $H\gtrsim H_{c1}$, $\zeta$ becomes approximately commensurate, $\zeta\approx 2/3$, at which point the first and second harmonics merge. Gaussian fits to the peak positions and widths of the observed peaks are plotted in Figs.~\ref{fig2}(d) and (e).  As one can see, there is hysteresis in the transition, which is consistent with a first-order commensuration transition in the magnetic order.

This transition can also be detected as a jump in magnetization when raising or lowering the field. This effect is most clearly seen in $M/H$, as shown in Fig.~\ref{fig2}(c).  The value of $H_{c1}$ is associated with the minimum in $M/H$.  Qualitatively similar behavior has been observed previously (but not explained) in EuAuSb \cite{ram2024magnetotransport,roy2024chiral}, EuAuAs \cite{malick2022electronic}, EuAgAs \cite{laha21}, EuCuSb \cite{takahashi2020competing}, and EuCuAs \cite{soh2024weyl}.  Our neutron results allow us to identify this as a field-induced commensuration transition of the spiral magnetic order.

Further insight into the character of the magnetic order is provided by analysis of a Heisenberg spin  Hamiltonian commonly used to describe helimagnetic order \cite{yoshimori1959_spiral,kaplan1959_spiral,villain1959_spiral,naga62,zali07}. A strong coupling between the Eu ions in the same layer describes the ferromagnetism.  Given that EuAuSb is semimetallic, we expect the exchange interactions between the layers to extend beyond nearest-neighbors, following the Ruderman-Kittel-Kasuya-Yosida model \cite{vanv62}; we consider effective antiferromagnetic exchange couplings between Eu sites in nearest-neighbor and next-nearest-neighbor layers, $J_1$ and $J_2$.  The dipolar interaction, with possible contribution from spin-orbit effects, leads to the observed easy-plane spin anisotropy, and the effect of the magnetic field is included through a spin Zeeman term.  In zero field, the magnetic wave vector $\zeta$ is given by the formula $\cos(\pi\zeta) = -J_1/4J_2$ \cite{naga62}.  To evaluate the effective exchange couplings, we performed calculations using density functional theory, with fixed spins, for four different collinear magnetic structures with unit cells containing 2, 4, or 6 Eu layers.  From those results, we estimate that $J_1=1.49$~meV and $J_2=0.57$~meV \footnote{Molecular field theory results for the Curie-Weiss temperatures and the saturation field provide an estimate of the sum of the exchange interactions \cite{john17}.  While inclusion of the ferromagnetic in-plane coupling partially offsets the antiferromagnetic interlayer couplings, the magnitudes of the calculated $J_1$ and $J_2$ values seem to be about an order of magnitude larger than the estimated scale; nevertheless, the ratio of values is consistent with the observed periodicity of the helical order.}, yielding a prediction of $\zeta = 0.727$, which is comparable to the low-temperature experimental value of 0.63.

The analysis of this model with an in-plane magnetic field \cite{zali95} indicates that the distortion of the spiral involves the appearance of finite second and higher harmonics.  The intensity of the second harmonic is predicted to be proportional to $H^2$ and to have a magnitude equal to 1/4 of the field-induced uniform response, as detected at the $L=2n$ Bragg peaks.  The first harmonic intensity should decrease, to leading order in $H$, as $1-\alpha H^2$, where $\alpha$ is a constant that depends on the exchange parameters.  As demonstrated in Fig.~\ref{fig2}(a), our measured intensities are in excellent agreement with these predictions for $H<H_{c1}$.

\begin{figure}
\includegraphics[width=\columnwidth]{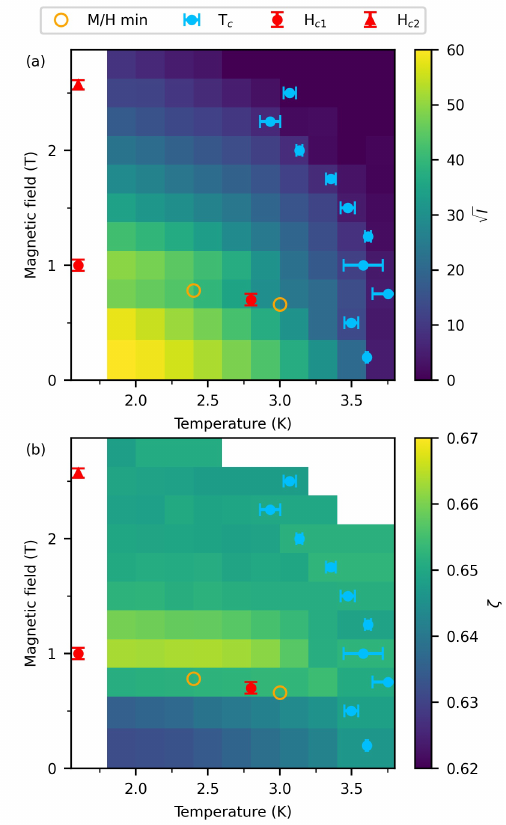}
\caption{(a) Color plot showing intensity of the magnetic $(0,0,2-\zeta)$ peak ($\sqrt{I}$ is plotted to enhance weak features), collected by fixing the magnetic field and sweeping the temperature. $T_c$ data points were extracted from this data set by fitting critical type behavior. The incommensurate to commensurate ($H_{c1}$) and high field transition ($H_{c2}$) values were extracted from the field-dependent data shown in Fig.~\ref{fig2}. The position of the minimum in the M/H data plotted in Fig.~\ref{fig2}(c) is also shown. (b) Color plot showing $\zeta$ extracted from the position of the magnetic $(0,0,2-\zeta)$ peak. }
\label{fig3}
\end{figure}

Detailed temperature-dependent measurements of the $(0,0,2-\zeta)$ peak were collected at constant field to map out the low temperature phase diagram. The peak intensity and position $\zeta$ are plotted in the colormaps displayed in Fig.~\ref{fig3}. The constant-field scans were fit with an order-parameter-like $T$-dependence $\sim(T_c-T)^{\alpha}$ to extract the critical temperature and these points are overlaid on the plots. We also extracted critical field information from the field-dependent data shown in Fig.~\ref{fig2}. The field values $H_{c2}$ at which the harmonic intensity disappears were found by fitting the $(0,0,2-\zeta)$ peak intensity with an order parameter behavior, and the $H_{c1}$ was chosen as the lowest field where $\zeta$ becomes fixed to its high-field value. We note that the precise value of $\zeta$ in the high field regime depends on the sample history (temperature or field cycling). This is likely due to differences in the concentration of defects in the helical order, such as phase slips where two adjacent layers are aligned ferromagnetically.

We have not yet considered the magnitude of the ordered magnetic moment.  The neutron diffraction intensities for magnetic peaks are proportional to the square of the component of the ordered moment oriented perpendicular to {\bf Q}.  We can write the integrated intensity for a peak centered at ${\bf Q}=(0,0,L)$ as
\begin{equation}
I_{\rm int}(L) = AC(L) f^2(L)|F(L)|^2,
\end{equation}
where $A$ is a scale factor, $C(L)$ is the factor describing absorption effects and the Lorentz factor, $f(L)$ is the Eu magnetic form factor, and $F(L)$ is the magnetic structure factor, normalized to the chemical unit cell.  At low temperature and zero field, full helical order with in-plane moments should give $|F(2-\zeta)|^2+|F(2+\zeta)|^2 = (2g\mu_B S)^2$ \footnote{We note that these two satellite peaks have equal structure factor. We compute the sum for more convenient comparison to $|F(2)|^2$, and to mitigate any inaccuracy in the form factor correction.}, assuming a spin-only $g$-factor of $g=2$. At in-plane field $H$, we should have $\Delta|F(2)|^2 = |F(2)|^2_H - |F(2)|^2_{H=0} = M^2$, where $M$ is the magnetization per unit cell, which should approach $2g\mu_B S$ at sufficiently high field. Hence, if we plot $|F|^2_{\rm tot}(H) = |F(2-\zeta)|^2+|F(2+\zeta)|^2 + \Delta|F(2)|^2$, it should give us a measure of the square of the average ordered in-plane spin component.

\begin{figure}[t]
\includegraphics[width=\columnwidth]{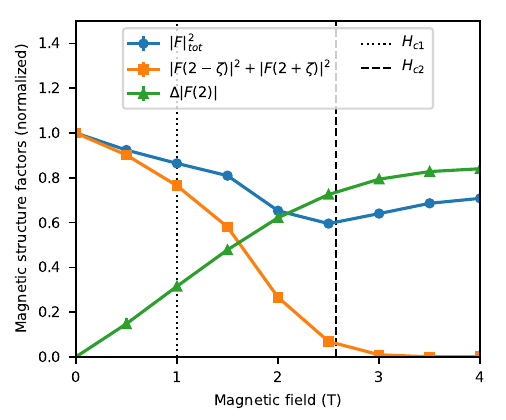}
\caption{Plot of $|F|^2_{\rm tot}$, $|F(2-\zeta)|^2+|F(2+\zeta)|^2$, and $\sqrt{\Delta|F(2)|^2}$ (designated $\Delta|F(2)|$ in the legend), normalized at $H=0$, as a function of magnetic field at $T=1.6$~K. }
\label{figmi}
\end{figure}

\begin{figure*}[t]
\includegraphics[width=1.7\columnwidth]{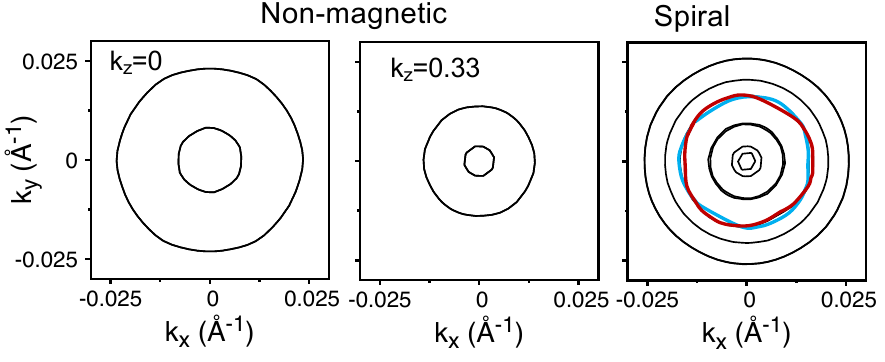}
\caption{Calculated Fermi surface intersections with the $k_x$-$k_y$ plane for fixed values of $k_z$, evaluated at a chemical potential ($-0.3$~eV) consistent with photoemission results. Left panels show results for the nonmagnetic state at $k_z=0$ and $k_z=0.33(2\pi/c)$. Right panel shows results at $k_z=0$ for commensurate helical order, where the larger real-space unit cell results in a reduced volume in reciprocal space, with bands at $k_z=\pm0.33(2\pi/c)$ folded back to $k_z=0$. Red and blue colors indicate opposite spin polarizations for states along the same radial {\bf k} vector. (Possible spin splitting of other Fermi pockets is too small to resolve in this figure.) }
\label{figfs}
\end{figure*}

We present $|F|^2_{\rm tot}(H)$ at $T=1.6$~K in Fig.~\ref{figmi}. We observe that it decreases to a minimum near $\mu_0H=2.5$~T, which is where the intensities of the satellite peaks approach zero.  The loss of intensity could, in principle, involve a static spin component perpendicular to the planes, to which the present measurements would be insensitive. We note that the previously discussed Heisenberg model predicts that field will induce a canted helical state with no such perpendicular component. However, recent studies of the helical order in layered YMn$_6$Sn$_6$ \cite{ghim20} and Fe$_3$Ga$_4$ \cite{bara25} have instead identified a field-induced transition to a transverse conical spiral (TCS) phase; a topological Hall effect (THE) is observed in the same temperature and field regime.  We argue that this is unlikely in our case since we do not see any large feature in the magnetization at $H_{c1}$, while the transition to the TCS phase involves a large jump in magnetization. A significant spin-flop component is therefore unlikely in EuAuSb; nevertheless, a topological Hall effect has been observed, with a broad maximum for $\mu_0H\sim2.5$~T \cite{yao25}, where the minimum occurs in $|F|^2_{\rm tot}(H)$. (Note that the THE observed in \cite{ram2024magnetotransport,roy2024chiral} was measured with field along $c$, which is different from our geometry.)

Analysis of the THE is based on the presence of noncoplanar orientations among neighboring spins \cite{naga13}. For the case of the TCS phase \cite{ghim20,bara25}, it was found necessary to invoke the participation of a helical magnon in order to explain the observed THE \cite{afsh21}.  The fact that the THE was observed to increase with temperature was taken to be consistent with a fluctuational THE.

In the present case, we are interested in $T = 1.6$~K and $\mu_0H=2.5$~T, for which $k_{\rm B}T/g\mu_{\rm B}H\approx0.07$, making thermal fluctuations irrelevant.  Instead, we suggest that the depression in $|F|^2_{\rm tot}(H)$ is due to quantum fluctuations induced by the in-plane magnetic field.  Note that it occurs where the sum of the intensities of the harmonic peaks is decaying to zero and before the uniform magnetization has reached saturation.  

To see the impact of the magnetic order on the electronic states near the Fermi level in zero field, we have performed electronic structure calculations for both the nonmagnetic and commensurate helical states, as described in Sec.~II. In Fig.~\ref{figfs}, we compare the 2D Fermi-surface (FS) plots for the non-magnetic and spiral magnetic phases at different $k_z$ values. The Fermi level is shifted by $\approx-$300 meV, in line with angle-resolved photoemission results \cite{yao25}; note that this shift is within the uncertainty of the calculations. 
In the nonmagnetic phase, two modulated $\Gamma$-centered pockets appear, giving rise to pockets of different sizes at $k_z=0$ and $k_z=\pm0.33$, in units of $2\pi/c$. (Note that, because of inversion symmetry, the FS is symmetric about $k_z=0$.) Furthermore, each pocket is spin-degenerate due to Kramers degeneracy, enforced by parity $\mathcal{P}$ and time-reversal $\mathcal{T}$ symmetries.
In the spiral phase, the FS becomes asymmetric about $k_z=0$ due to the broken inversion symmetry, so that when the pockets at $k_z=\pm0.33$ are folded back to the $k_z=0$ plane, we get 6 pockets. Moreover, additional spin splitting appears as $\mathcal{PT}$ symmetry is no longer preserved; the magnitude of such spin-splitting varies among the bands. We note that this splitting shows the expected symmetry for the $6’22’$ point group of the commensurate magnetic structure, and is also non-relativistic: it is observed even when \gls*{SOC} is set to zero (unlike relativistic Rashba splitting which relies on the presence of \gls*{SOC}).

\section{Discussion}

In the low-temperature phase for $H<H_{c1}$, the incommensurate wave vector, with $\zeta=2/3 - \delta$, is determined by the indirect exchange between Eu sites via states involving orbitals on intervening Au and Sb sites. For small but finite $\delta$, the incommensurate spiral can be viewed as a long-wavelength modulation (rotation) of an ideal triangular structure with a three-layer period (i.e., the commensurate spin spiral where spins rotate by 120° from one layer to the next).  If the spiral were commensurate, with $\delta=0$, application of an in-plane magnetic field would cause the spin triad to rotate to a unique energy-minimizing orientation with every third spin antialigned to the field direction \cite{Chubukov_1991}. For non-zero $\delta$, local twists can improve the alignment of those spins close to the field direction \footnote{See case A in Sec.~V of Ref. [34]}. These solitonic distortions, reflected in the development of higher harmonics \cite{zali95}, grow as the in-plane field is increased; $\delta$ (and $\zeta$) also begin to change.
Before the change in $\zeta$ can grow too large, we observe a first-order transition to a state that is approximately commensurate. 

For $H>H_{c1}$, the commensurate structure can then take on the low-energy configuration of Ref. \cite{Chubukov_1991}. This compromise state is likely to come at a cost of increased resistivity, as suggested by previous work \cite{ram2024magnetotransport,roy2024chiral}. With increasing field we find that the diffracted intensity from the helical order decreases, while the ferromagnetically-polarized component detected at (0,0,2) grows and eventually saturates. Taken together, this state appears to involve a canted helix, with the in-plane spins canting toward the field direction.  The deviations from commensurability with increasing field and especially temperature are consistent with discommensurations \cite{sosi04} or solitons \cite{toga16} in which pairs of neighboring layers are oriented along the applied field. We note that our findings are consistent with quantum theory for triangular magnets \cite{Chubukov_1991}, which predicts a transition to an up-up-down structure where two out of three spins are aligned with magnetic field at $H = H_{sat}/3$, which is roughly consistent with the commensuration field $H_{c1}$ we observe. Above $H_{sat}/3$, a canted structure is predicted where spins progressively rotate towards the field direction.

The helical order is chiral, but, in zero field, right-handed and left-handed spirals should be equivalent and equal domains of both should be expected.  Cooling through $T_N$ while applying electric current in combination with magnetic field parallel to the $c$ axis may preferentially select one domain \cite{Jiang2020_helimagnet, Masuda2024_helimagnet}, which should lead to detectable effects in transport measurements.

One surprising feature is that the value of $T_N$ indicated by the magnetic susceptibility, 3.9~K, is greater than that of 3.5~K indicated by the neutron results in Fig.~\ref{fig3}(a). If the magnetic moments were to rotate to be parallel to the $c$ axis, as observed in EuCuSb \cite{takahashi2020competing}, the intensity of magnetic peaks at $(0,0,2n\pm\zeta)$ would be zero; however, the temperature dependence of the anisotropy in the magnetic susceptibility data is inconsistent with such behavior.  Another possibility is that the interlayer magnetic coupling is effectively frustrated, so that the initial order is two-dimensional and short-ranged, followed by long-range three-dimensional order at a lower temperature.  This possibility will be considered further in \cite{yao25}.

Lastly, we discuss our electronic structure calculations which show that the observed helical order induces spin-splitting in the electronic bands due to the breaking of parity and time-reversal symmetries. Since the magnetic moments are also fully compensated it is natural to draw comparisons to the recently recognized collinear altermagnets, which show spin-splitting that is even under inversion of reciprocal space. In contrast, the splitting observed in EuAuSb and plotted in Fig.~\ref{figfs} has odd-wave parity. A number of recent studies have explored materials showing a similar odd-wave symmetry in their band structure (sometimes termed $p$-wave or $f$-wave depending on symmetry) \cite{hellenes2024pwavemagnets, pari25, song2025pwaveswitch, kudasov2024helical, cuono2023_eucompounds}, and we therefore identify EuAuSb as a member of this class of materials with non-relativistic spin-splitting arising from the magnetic symmetry.

\section{Conclusions}

In summary, we have measured magnetic Bragg peaks in a single crystal of EuAuSb and identified magnetic order with a magnetic unit cell containing approximately three layers of Eu atoms. Our observations are consistent with ferromagnetic ordering within the layers, with the net moment rotating by $\sim$120$^{\circ}$ from one layer to the next. Furthermore, we identify the metamagnetic transition occurring near 1~T as an incommensurate to commensurate transition where the ordering wave vector locks in to $Q=(0,0,\frac{2}{3})$. An analogous transition should occur in a number of other Eu$MX$ compounds and has been seen in EuCuAs \cite{soh2024weyl}.  The ordering wave vector is consistent with a model of competing antiferromagnetic interlayer couplings, and the evolution of the incommensurate order with an in-plane field follows the predictions for a helical magnet with easy-plane anisotropy \cite{zali95}.  The depression in sum of squares of static in-plane spin components at the field where magnetic harmonics disappear correlates with an observation of the topological Hall effect; we propose that both are associated with quantum spin fluctuations.  The magnetic order also has a feedback effect on the electronic states, with the resulting effective magnetic field in each AuSb layer causing an odd-parity spin splitting of the Dirac-cone states with 3-fold symmetry, which should have interesting consequences for transport measurements. 

\acknowledgments

JMT and IZ acknowledge helpful discussions with I. I. Mazin, and JS and JMT acknowledge insightful comments from R. M. Fernandes. Work at Brookhaven is supported by the Office of Basic Energy Sciences, Materials Sciences and Engineering Division, U.S.\ Department of Energy (DOE) under Contract No.\ DE-SC0012704. This research used resources at the High Flux Isotope Reactor, a DOE Office of Science User Facility operated by Oak Ridge National Laboratory. 

\section*{Data Availability}
The data that support the findings of this article are openly available on the Zenodo database under the access code 17048126 \cite{repo}.

\end{document}